\documentclass[twocolumn,prb,showpacs]{revtex4}
\begin{document}
\topmargin-1.0cm

\title {Intersubband spin-density excitations in quantum wells with Rashba spin splitting}
\author {C. A. Ullrich}
\affiliation {Department of Physics,  University of
Missouri-Rolla, Rolla, Missouri 65409}
\author {M. E. Flatt\'e}
\affiliation {Department of Physics and Astronomy, University of Iowa,
Iowa City, Iowa 52242}
\date{\today}

\begin{abstract}

In inversion-asymmetric semiconductors, spin-orbit coupling induces a
{\bf k}-dependent spin splitting of valence and conduction bands, 
which is a well-known cause for spin decoherence in bulk and heterostructures.  
Manipulating nonequilibrium spin
coherence in device applications thus requires understanding how 
valence and conduction band spin splitting affects carrier spin dynamics.
This paper studies the relevance of 
this decoherence mechanism for collective intersubband spin-density excitations (SDEs) in quantum wells. 
A density-functional formalism for the linear spin-density matrix response is presented that 
describes SDEs in the conduction band of quantum wells with subbands that may be non-parabolic and spin-split
due to bulk or structural inversion asymmetry (Rashba effect).
 As an example, we consider a 40 nm GaAs/Al$_{0.3}$Ga$_{0.7}$As quantum well,
including Rashba spin splitting of the conduction subbands. We find a coupling and wavevector-dependent splitting of
the longitudinal and transverse SDEs. However, decoherence of the SDEs is not determined by 
subband spin splitting, due to collective effects arising from dynamical exchange and correlation.
\end{abstract}

\maketitle

\newcommand{\scy}{\rm \scriptscriptstyle}
\newcommand{\exc}{e_{\rm xc}^h}
\newcommand{\bia}{^{\rm \scriptscriptstyle BIA}}
\newcommand{\sia}{^{\rm \scriptscriptstyle SIA}}
\newcommand{\alda}{^{\rm \scriptscriptstyle ALDA}}
\newcommand{\dy}{\displaystyle}
\newcommand{\nuu}{n_{\uparrow \uparrow}}
\newcommand{\nud}{n_{\uparrow \downarrow}}
\newcommand{\ndu}{n_{\downarrow \uparrow}}
\newcommand{\ndd}{n_{\downarrow \downarrow}}
\newcommand{\qp}{q_{||}}
\newcommand{\rp}{r_{||}}
\newcommand{\kp}{k_{||}}
\newcommand{\qqp}{{\bf q}_{||}}
\newcommand{\rrp}{{\bf r}_{||}}
\newcommand{\kkp}{{\bf k}_{||}}
\newcommand{\ua}{\uparrow}
\newcommand{\da}{\downarrow}
\newcommand{\tdlda}{^{\scriptscriptstyle \rm TDLDA}}
\newcommand{\ks}{^{\scriptscriptstyle \rm KS}}
%******************************************************************************
\section{introduction} \label{sec:intro}

Most currently available semiconductor device technologies are entirely based
on manipulating electronic charges. The emerging field of spintronics \cite{awschalombook,wolf}
proposes to exploit, in addition, the spin degree of freedom of carriers, thereby 
adding new features and functionalities to solid-state devices. 
Many of the proposed new applications rely, in one form or another, 
on manipulating nonequilibrium spin coherence.
The hope that this may indeed lead to  viable practical approaches
has been supported through recent experimental observations 
\cite{kikkawa1,kikkawa2,kikkawa3} of long-lived ($>100$ ns) and
spatially extended ($>100$ $\mu$m) coherent spin states in semiconductors.
Two characteristic times, $T_1$ and $T_2$, provide a quantitative measure for the 
magnitude and persistence of spin coherence. $T_1$ describes the return to equilibrium of 
a non-equilibrium spin population, and $T_2$ measures the coherence loss
due to dephasing of transverse spin order (for more details, see Ref. \onlinecite{flatte}). 

Spin relaxation in GaAs quantum wells was recently studied experimentally \cite{wagner,terauchi,tackeuchi,ohno,sandhu}
and theoretically. \cite{DK,averkiev,bournel,lau,lau1}
Measurements of the electronic $T_1$ involve circularly polarized pump-probe
techniques to create and observe coherent spin populations in the lowest conduction subband. 
Electron spin decoherence has been shown to occur via the spin precession \cite{DP} of
carriers with finite crystal momentum $\bf k$ in the effective $\bf k$-dependent crystal magnetic
field of an inversion-asymmetric material. \cite{dresselhaus} We note that the theory 
of Refs. \onlinecite{flatte,lau} gives good agreement with experiment, 
without including electron-electron interactions. 

In this paper, we will consider electronic charge and spin dynamics in quantum wells 
involving not {\em one} but {\em two} subbands (we will limit the discussion here mainly
to conduction subbands). One motivation for this work is that intersubband (ISB) 
{\em charge} dynamics in quantum
wells is currently of great experimental and theoretical interest, \cite{books}
since electronic ISB transitions are the basis of a variety of new devices operating in the terahertz
frequency regime, such as detectors, \cite{detector} modulators \cite{modulator}
and quantum cascade lasers. \cite{laser,kohler}
In view of this, it seems worthwhile to explore ISB {\em spin} dynamics as a possible
route towards novel applications in the terahertz regime. 

Analogous to the case of spin dynamics discussed above, one may define characteristic
times for {\em intersubband} dynamics (for an overview, see Ref. \onlinecite{helm}). Population decay
from an excited to a lower conduction subband is measured by an intersubband relaxation time 
$T_1^{\scy ISB}$, and loss of coherence of collective intersubband excitations is
measured by a dephasing time $T_2^{\scy ISB}$. These two times have been measured
experimentally for ISB charge-density excitations in quantum wells \cite{heyman,williams} 
and found to differ substantially at low
temperatures, $T_2^{\scy ISB}$ being three orders of magnitude smaller than $T_1^{\scy ISB}$.
The reason is that intersubband relaxation proceeds mainly via phonon emission 
and is thus slowed down by an energy bottleneck for small phonon momenta as well as for the optical
phonon branch. 
This differs from the case of conduction electron spin relaxation,
where $T_1$ and $T_2$ are comparable. \cite{flatte,lau}

On the other hand, dephasing of collective ISB excitations in 
quantum wells is determined by a complex interplay of a variety of different scattering
mechanisms, whose relative importance is not {\em a priori} obvious. 
In recent experimental \cite{jon} and theoretical \cite{PRL} work, it was found
that the linewidth of (homogeneously broadened) ISB charge plasmons in a wide 
GaAs/Al$_{0.3}$Ga$_{0.7}$As quantum well, where phonon scattering plays no role,
is determined mainly by interface roughness and electronic many-body effects. 

The question now arises which physical mechanisms govern the dephasing of {\em collective} ISB
spin-density excitations. As a first step towards a clarification of this question,
this paper addresses the influence of the $\bf k$-dependent crystal magnetic field in 
semiconductor quantum wells on ISB spin-density excitations, and the importance of
many-body effects. The latter will be described in the framework of density-functional theory
(DFT).

This paper is organized as follows. In Section \ref{sec:groundstate}
we set up the formalism for calculating the electronic ground state in modulation-doped 
quantum well conduction subbands, including spin-orbit coupling and many-body effects.
Section \ref{sec:response} presents a general response formalism for the 
spin-density matrix, based on time-dependent DFT (TDDFT). 
In Section \ref{sec:results} we consider an explicit example
and calculate the collective ISB charge- and spin-density excitations in the conduction band of a 
GaAs/Al$_{0.3}$Ga$_{0.7}$As quantum well, including spin-orbit coupling.
Section \ref{sec:conclusion} contains our conclusions.
Various technical details can be found in Appendices \ref{appendixA} and \ref{appendixB}.

%*******************************************************************************
\section{Electronic ground state}\label{sec:groundstate}

We consider a modulation-doped quantum well (direction of growth: $z$-axis)
containing $N$ conduction electrons per unit area.
In the standard multi-band $\bf k \cdot p$ approach for semi\-conductors,
\cite{kane,bastard,yu,vasko} the single-particle states in a quantum well are 
expanded in terms of Bloch functions at the zone center, $u_n({\bf r})$:
\begin{equation} \label{II.1}
\Psi_{j\qqp}({\bf r}) = \sum_{n=1}^{N_{b}}
e^{i {\bf q}_{||} {\bf r}_{||}} \: \psi_{jn}(z) \: u_n({\bf r})\;,
\end{equation}
where $\psi_{jn}(z)$ are envelope function belonging to the $j$th subband,
and ${\bf r}_{||} = (x,y)$ and ${\bf q}_{||} = (q_x,q_y)$ are in-plane position and wave vectors.
In general, $u_n({\bf r})
= u_{n\uparrow}({\bf r}) \xi_\uparrow + u_{n\downarrow}({\bf r}) \xi_\downarrow$,
where $\xi_{\uparrow,\downarrow}$ denote two-component Pauli spinors.
The functions $u_{n\sigma}({\bf r})$ have the form
$u_{n\sigma}({\bf r}) = c_{n\sigma} |j\rangle$, where $c_{n\sigma}$ are  complex
coefficients. 
Usually $N_b$ includes at least the conduction-band $s$ states and the valence-band $p$ states
(8-band or Kane \cite{kane} model), but in general a 14-band model is needed for a consistent
description of spin splitting in heterostructures. \cite{rossler} In that case, the set of basis functions
is 
\begin{equation} \label{II.2}
|j\rangle \in \Big\{ |Z\rangle_v , |X \pm iY\rangle_v, |S\rangle_c,
|Z\rangle_c, |X \pm iY\rangle_c \Big\} \;.
\end{equation}
This leads to a Hamiltonian in $8\times 8$ (or $14\times 14$) matrix form, whose elements
are well known. \cite{bastard,yu} The envelope functions $\psi_{jn}(z)$ for valence and
conduction bands then follow from the resulting 8 (or 14) coupled single-particle
equations. \cite{bastard} 

If one is only interested in the electronic structure of the
conduction band of a quantum well, it is convenient to reduce the multi-band Hamiltonian
described above to a $2\times 2$ conduction band Hamiltonian.
\cite{rossler,warburton,eppenga,andrada,pfeffer} 
The single-particle states (\ref{II.1}) can then be simplified to the following two-component form:
\begin{equation} \label{II.3}
\Psi_{j{\bf q}_{||}}({\bf r}) = e^{i \qqp \rrp} \left( \begin{array}{c}
\varphi_{j\uparrow}(\qqp,z) \\ \varphi_{j\downarrow}(\qqp,z) \end{array} \right).
\end{equation}
The envelope functions $\varphi_{j\sigma}$ follow from a two-component
effective-mass Kohn-Sham equation: 
\begin{eqnarray}  \label{II.4}
\lefteqn{ \hspace{-2.5cm}
\sum_{\beta = \uparrow,\downarrow} \left( 
\hat{h}\:\delta_{\alpha \beta} 
+ v^{\rm ext}_{\alpha \beta}(z) 
+ \hat{H}^{\rm so}_{\alpha \beta}(z) 
+ v^{\rm xc}_{\alpha \beta}(z) \right) \varphi_{j\beta} (\qqp,z)} \nonumber\\
&=& E_{j\qqp} \,\varphi_{j\alpha} (\qqp,z)\;, 
\end{eqnarray}
where $\alpha = \uparrow,\downarrow$, and
\begin{equation}\label{II.5}
\hat{h} = - \, \frac{d}{ dz} \, \frac{\hbar^2}{2 m(E,z)} \, \frac{d}{dz}
+ \frac{\hbar^2 q_{||}^2}{2 m(E,z)} +v_{\rm conf}(z) + v_{\rm H}(z) .
\end{equation}
The spin-independent part $\hat{h}$ of the $2\times 2$ conduction band Hamiltonian
accounts for possible non-parabolicity of the subbands through an
effective mass that depends on $E_{j\qqp}$. Explicit expressions for $m(E,z)$ 
can be found in Refs. \onlinecite{andrada,pfeffer}. 
$v_{\rm conf}(z)$ is the confining bare quantum well potential (e.g., a square well),
and the Hartree potential $v_{\rm H}(z)$  is related to the electron ground-state
density $n(z)$, defined below, through Poisson's equation:
\begin{equation} \label{II.6}
\frac{d^2 v_{\rm H}(z)}{dz^2} = - 4\pi {e^*}^2 n(z) \;, 
\end{equation}
where $e^* = e/\sqrt{\epsilon}$ is the effective charge ($\epsilon$ is the static
dielectric constant of the material).

Let us now discuss the spin-dependent parts of the Hamiltonian in Eq. (\ref{II.4}).
The first term, $v^{\rm ext}_{\alpha \beta}(z)$, describes externally applied
uniform static electric and magnetic fields,
${\bf E}$ and $\bf B$:
\begin{equation} \label{II.7}
v^{\rm ext}_{\alpha \beta}(z)=
e E_z z \: \delta_{\alpha \beta}
+ \frac{1}{2} \: g^*(z) \mu_B \: {\bf B} \cdot \vec{\sigma} \;,
\end{equation}
where ${\bf E}= \hat{e}_z E_z$ is perpendicular to the quantum well, and $\bf B$ can have arbitrary direction.
$\vec{\sigma}$ is the vector of the Pauli spin matrices, and $g^*(z)$ denotes the g-factor of the bulk
material at point $z$. 

Intrinsic conduction band spin splitting, caused by  spin-orbit interaction, in general comes from several
different sources. One often deals with situation where there are
two major contributions:  $\hat{H}^{\rm so}_{\alpha \beta}=
\hat{H}\bia_{\alpha \beta}+\hat{H}\sia_{\alpha \beta}$, where BIA and SIA denote bulk and structural inversion
asymmetry. The first term has the well-known form 
$\hat{H}\bia_{\alpha\beta}=[\hbar {\bf \Omega }\cdot \vec{\sigma}/2]_{\alpha\beta}$,
where ${\bf \Omega} = \gamma \Big(q_x(q_y^2-q_z^2),q_y(q_z^2-q_x^2),q_z(q_x^2-q_y^2)\Big)$
for bulk zincblende semiconductors. \cite{dresselhaus}
For a quantum well, $\hat{H}\bia_{\alpha \beta}$
depends on the growth direction. For instance, along [001] we have 
\begin{eqnarray} \label{II.12}
\hat{H}\bia_{\uparrow\uparrow}=\left(\hat{H}\bia_{\downarrow\downarrow}\right)^\dagger
&=&i(q_x^2-q_y^2)\left(\frac{1}{2} \frac{d \gamma}{dz} + \gamma \frac{d}{dz}\right)
\nonumber\\
\hat{H}\bia_{\uparrow\downarrow}=\left(\hat{H}\bia_{\downarrow\uparrow}\right)^\dagger
&=& 
 - \frac{d}{dz}\,\gamma \,\frac{d}{dz}(q_x + i q_y)\nonumber\\
 && {} - i\gamma q_xq_y(q_x-iq_y) \;,
\end{eqnarray}
for [110] and [111] directions, see Ref. \onlinecite{eppenga}.
The second contribution to intrinsic spin splitting, SIA (also known as the Rashba effect \cite{rashba}),
has the form 
\begin{eqnarray} \label{II.13}
\hat{H}\sia_{\uparrow\uparrow} &=& \hat{H}\sia_{\downarrow\downarrow} = 0 \\
\hat{H}\sia_{\uparrow\downarrow} &=& \left(\hat{H}\sia_{\downarrow\uparrow}\right)^\dagger
= -\frac{i}{2} \frac{d\eta}{dz} \: (q_x - i q_y) \;. \label{II.14}
\end{eqnarray}
The material parameters $\gamma(z)$ and $\eta(z)$ are explicitly given in Ref. \onlinecite{pfeffer}.
We also mention a possible additional source of spin splitting in quantum wells,
the so-called native interface asymmetry, related to chemical bonds across interfaces (for more
details see Refs. \onlinecite{flatte,NIA}).

A novel feature in Eq. (\ref{II.4}), which distinguishes the present approach
from previous studies of conduction band non-parabolicity and spin splitting,
is that many-body effects are explicitly included through the 
exchange-correlation (xc) potential $v^{\rm xc}_{\alpha\beta}(z)$.
xc effects have previously been shown to produce non-negligible shifts of
quantum well subband energies and ISB charge plasmon
frequencies. \cite{helm,PRL,quantumwell} As discussed below, including xc effects is crucial
for a physically correct description of collective ISB spin excitations.

The solutions of Eq. (\ref{II.4}) have the 
interesting property of being mixed spin-up and spin-down eigenstates,
due to the off-diagonal terms in the Hamiltonian caused by spin-orbit coupling and, 
possibly, externally applied transverse magnetic fields. The off-diagonal terms
in $\hat{H}^{\rm so}_{\alpha\beta}$
depend on $\qqp$, and there is no choice of basis which diagonalizes 
Eq. (\ref{II.4}) for {\em all} $\qqp$. Due to the absence of a global quantization
axis, spin is no longer a good quantum number. This requires a generalization of the
well-known spin-DFT \cite{gunnarsson} to systems with non-collinear spin. 
So far, this was done at only few occasions in the literature, namely for non-collinear
magnetic materials such as $\gamma$-Fe, U$_3$Pt$_4$ and Mn$_3$Sn, \cite{kubler,sandratskii}
and inhomogeneous quantum Hall systems, \cite{heinonen} but, to our knowledge,
never before in the present context of semiconductor nanostructures.

Formally, the xc potential is defined as
\begin{equation} \label{II.15}
v^{\rm xc}_{\alpha \beta}({\bf r}) = \frac{ \delta E_{\rm xc}[\, \underline{n}\, ]}{\delta
n_{\alpha \beta}({\bf r})} \;,
\end{equation}
where the xc energy of the system, $E_{\rm xc}[\, \underline{n}\, ]$,
is a functional of the spin-density matrix \cite{gunnarsson}
\begin{equation}\label{II.16}
\underline{n}({\bf r}) = \sum_{j,\qqp} f_{j\qqp} \Psi_{j \qqp} \Psi^\dagger_{j \qqp}
\equiv
 \left( \begin{array}{cc} \nuu & \nud\\ \ndu & \ndd \end{array} \right),
\end{equation}
where $f_{j \qqp} \equiv f(E_F - E_{j\qqp})$ denotes the Fermi occupation function,
and $E_F$ is the conduction band Fermi level.
For $\Psi_{j \qqp}$ given by Eq. (\ref{II.3}), we have
\begin{equation}
\nuu(z) = \sum_{j,\qqp} f_{j\qqp}|\varphi_{j \uparrow}(\qqp,z)|^2 
\end{equation}
[similarly for $\ndd(z)$], and
\begin{equation}
\nud(z) = \ndu^*(z) = \sum_{j,\qqp} f_{j\qqp}\varphi_{j \uparrow}(\qqp,z) \varphi_{j \downarrow}^*(\qqp,z)\; .
\end{equation}
The usual approximation is to take the 
density matrix (\ref{II.16}) to be locally diagonal, \cite{sandratskii,heinonen}
so that the LSDA for non-collinear spin reads
\begin{equation} \label{II.17}
v^{\rm xc}_{\alpha\beta}(z) = 
\left. \frac{\partial}{\partial n_{\alpha \beta}} \, [ n e_{\rm xc}^{\rm hom}(n,|\vec{\xi}|)]
\right|_{\underline{n} = \underline{n}(z)} .
\end{equation}
$e_{\rm xc}^{\rm hom}(n,\xi)$ is the xc energy per particle of a homogeneous electron gas of density
$n$ and spin polarization $\xi$, which is well known from quantum Monte Carlo calculations. \cite{VWN} 
The local density and spin polarization are given by
\begin{eqnarray} \label{II.18}
n &=& \mbox{Tr} \: \underline{n} \\
\vec{\xi} &=& \frac{1}{n} \: \mbox{Tr} \: \vec{\sigma} \: \underline{n} \:.
\label{II.19}
\end{eqnarray}
The ground-state density is normalized as $\int dz \: n(z) = N$.
Explicit expressions for $v^{\rm xc}_{\alpha \beta}(z)$ are given in Appendix \ref{appendixA}.
With this form for $v^{\rm xc}_{\alpha \beta}(z)$, the $2\times 2$ effective-mass
Kohn-Sham equation (\ref{II.4}) is now completely defined. Self-consistent solution yields
a set of subbands which are occupied up to  $E_F$.

%**********************************************************************************
\section{linear response formalism for the spin-density matrix}\label{sec:response}

Once the electronic ground state (characterized by a set of subband levels and wave functions) 
has been calculated, the next step is to consider excitations. 
The formal framework for describing excitations in electronic many-body systems is 
provided by {\em linear response theory}. \cite{grosskohn,petersilka}

For the case where the wave functions take on a two-component form,
the TDDFT linear response equation for quantum wells becomes a $2\times 2$ matrix equation:
\begin{eqnarray} \label{III.5}
n^{(1)}_{\sigma \sigma'} (\kkp,z,\omega) &=& \sum_{\lambda,\lambda'=\uparrow,\downarrow}
\int \! dz' \: \chi_{\sigma \sigma', \lambda \lambda'}^{\rm \scriptscriptstyle KS} (\kkp,z,z',\omega)
\nonumber\\
&& \times v^{\rm (1)}_{\lambda \lambda'} (\kkp,z',\omega) \;.
\end{eqnarray}
This expresses, formally exactly, the first-order change of the spin-density matrix, $n^{(1)}_{\sigma \sigma'}$,
via the response of a non-interacting
system, characterized by the response function $\chi_{\sigma \sigma', \lambda \lambda'}^{\rm \scriptscriptstyle KS}$
(see below), to an effective perturbing potential of the form
$v^{\rm (1)}_{\lambda \lambda'}   =
v^{(\rm 1,ext)}_{\lambda \lambda'}  +
v^{(\rm 1,H)}_{\lambda \lambda'} +
v^{(\rm 1,xc)}_{\lambda \lambda'} $.
Here, $v^{(\rm 1,ext)}_{\lambda \lambda'}$ is the external perturbation, and the linearized Hartree
and xc potentials are 
\begin{widetext}
\begin{eqnarray} \label{III.6}
\lefteqn{
v^{(\rm 1,H)}_{\lambda \lambda'}(\kkp,z',\omega) +v^{(\rm 1,xc)}_{\lambda \lambda'}(\kkp,z',\omega)}
\nonumber\\[2mm]
&=&
\sum_{\zeta,\zeta' = \uparrow,\downarrow} 
\int\!dz'' \left[ \frac{2 \pi {e^*}^2}{\kp}\: e^{-\kp|z'-z''|} \:\delta_{\lambda \lambda'}\delta_{\zeta \zeta'} +
 f^{\rm xc}_{\lambda \lambda',\zeta \zeta'}
(\kkp,z',z'',\omega)\right] n^{(1)}_{\zeta \zeta'}(\kkp,z'',\omega) \;.
\end{eqnarray}
In the widely used adiabatic local-density approximation (ALDA), \cite{zangwill} the xc kernel is given by
\begin{equation} \label{III.7}
f^{\rm xc}_{\lambda \lambda',\zeta\zeta'}(\kkp,z,z',\omega) =
\frac{\partial^2 e_{\rm xc}^{\rm hom}(n,|\vec{\xi}|)}
{\partial n_{\lambda \lambda'}(z)\partial n_{\zeta \zeta'}(z)}
 \: \delta(z-z') \;.
\end{equation}
The non-interacting response function  takes on the form of a fourth-rank tensor:
\begin{equation} \label{III.8}
\underline{\underline{\chi}}^{\rm \scriptscriptstyle KS}({\bf r},{\bf r}',\omega) = 
\sum_{\scriptstyle jl }^{\infty} \sum_{\qqp\qqp'} (f_{j\qqp} - f_{l\qqp'}) \:
\frac{ \Psi_{j\qqp }({\bf r})\Psi^\dagger_{l\qqp'}({\bf r})
\Psi^\dagger_{j\qqp }({\bf r}')\Psi_{l\qqp'}({\bf r}')}
{\omega - E_{j\qqp} + E_{l\qqp'} + i\eta} \;,
\end{equation}
where the $\Psi$'s are given by Eq. (\ref{II.1}). Formally, this multiband response formalism 
describes transitions among valence and conduction subbands, 
as well as  interband transitions. 
In this paper, however, we focus exclusively on intersubband transitions 
in the conduction band of modulation doped heterostructures.  Using Eq. (\ref{II.3}),
one can then transform the response function (\ref{III.8}) into
\begin{eqnarray}  \label{III.9}
\chi^{\rm \scriptscriptstyle KS}_{\sigma\sigma',\lambda\lambda'}(\kkp,z,z',\omega) &=&
 \sum_{jl}\! \int \! \frac{d^2 \qp}{(2\pi)^2} \:
\frac{f_{l \qqp-\kkp}- f_{j\qqp}}
{\omega - E_{j\qqp} + E_{l\qqp-\kkp}+ i\eta}\nonumber\\[1mm]
&&\times  
\varphi_{j\sigma}(\qqp,z) \varphi^*_{l\sigma'}(\qqp-\kkp,z)
\varphi^*_{j\lambda}(\qqp,z') \varphi_{l\lambda'}(\qqp-\kkp,z') \:,
\end{eqnarray}
\end{widetext}
where the Kohn-Sham envelope functions $\varphi_{j\sigma}$ and energies
$E_{j \qqp}$ are obtained from Eq. (\ref{II.4}).

One can combine the perturbing spin-dependent potentials $v^{(1)}_{\sigma \sigma'}$ and the 
solutions $n^{(1)}_{\sigma\sigma'}$ of the response equation (\ref{III.5}) in the
following, physically more transparent way, see also Eqs. (\ref{a3}) and (\ref{a4}) of Appendix \ref{appendixA}:
\begin{equation} \label{III.10}
V_j^{(1)} = {\rm Tr} \left[\sigma_j \: \underline{v}^{(1)} \right]
\end{equation}
\begin{equation} \label{III.11}
m_j^{(1)} = {\rm Tr} \left[\sigma_j \: \underline{n}^{(1)} \right] \;,
\end{equation}
$j=0,1,2,3$, where $\sigma_0$ is the $2\times 2$ unit matrix, and 
$\sigma_1,\sigma_2,\sigma_3$ are the Pauli matrices.

$m_0^{(1)}=n^{(1)}_{\uparrow \uparrow} + n^{(1)}_{\downarrow \downarrow} $ 
describes a collective charge-density excitation (CDE),  and
$m_3^{(1)}=n^{(1)}_{\uparrow \uparrow} - n^{(1)}_{\downarrow \downarrow}$ is a 
longitudinal spin-density excitation (SDE) with respect to the $z$-axis. 
In terms of this choice of global spin quantization, 
$m_1^{(1)}=n^{(1)}_{\uparrow \downarrow} + n^{(1)}_{\downarrow \uparrow} $ 
and $m_2^{(1)}=i[n^{(1)}_{\uparrow \downarrow} - n^{(1)}_{\downarrow \uparrow}]$ appear as transverse
spin-density (or spin-flip) excitations.
The CDE couples to an oscillating electric 
field polarized along the $z$-direction, associated with $V_0^{(1)}$.
The longitudinal SDE is excited by an oscillating magnetic field along $z$ 
associated with $V_3^{(1)}$,
and the transverse SDEs are excited by oscillating magnetic fields along $x$ and $y$, associated with
$V_1^{(1)}$ and $V_2^{(1)}$, respectively. We will discuss these selection rules in more detail below.

In terms of these quantities, the linear response equation (\ref{III.5}) takes on the following form:
\begin{equation}\label{III.12}
m_j^{(1)}(\kkp,z,\omega) = \sum_{k=0}^3 \int \! dz' \: \Pi^{\rm \scriptscriptstyle KS}_{jk}
(\kkp,z,z',\omega)  V_k^{(1)}(\kkp,z',\omega) .
\end{equation}
The response functions $\Pi\ks_{jk}$ and $\chi\ks_{\sigma\sigma',\lambda\lambda'}$ 
are related as follows:
\begin{equation} \label{III.13}
\Pi^{\rm \scriptscriptstyle KS}_{jk} = \sum_{\sigma ,\sigma' \atop \lambda,\lambda'}
\chi^{\rm \scriptscriptstyle KS}_{\sigma \sigma',\lambda \lambda'}
\left( \frac{\partial m_j^{(1)}}{\partial n^{(1)}_{\sigma \sigma'}} \right)
\left( \frac{ \partial V_k^{(1)}}{\partial v^{(1)}_{\lambda \lambda'}} \right)^{-1},
\end{equation}
where the coefficients $\partial m_j^{(1)}/\partial n^{(1)}_{\sigma \sigma'}
\left[= \partial V_j^{(1)}/\partial v^{(1)}_{\sigma \sigma'}\right]$ 
are easily obtained  from Eqs. (\ref{III.10}) or (\ref{III.11}). 
Explicit expressions for $\Pi\ks_{jk}$  are given in Appendix \ref{appendixB}.
The $V_k^{(1)}$, in turn, are given as sums of external perturbations and linearized
Hartree and xc terms:
\begin{eqnarray} \label{III.14}
V_k^{(1)}(\kkp,z,\omega) &=& V^{\rm ext}_k(\kkp,z,\omega)
\nonumber\\
&+&
\sum_{l=0}^3 
\int\!dz' \left[ \frac{2 \pi {e^*}^2}{\kp}\: e^{-\kp|z-z'|} \:\delta_{k0}\delta_{l0} \right.
\nonumber\\
&+&
 f^{\rm xc}_{kl}(\kkp,z,z',\omega)\bigg] m_l^{(1)}(\kkp,z',\omega) .
\end{eqnarray}
The xc kernels $f^{\rm xc}_{kl}$ in ALDA are given in Appendix \ref{appendixA}.

%**********************************************************************************

\section{results and discussion}\label{sec:results}

\subsection{Kohn-Sham wavefunctions and Rashba effect}

We will now discuss an example to illustrate the spin-density matrix response formalism
developed above. Consider the case of a 40 nm wide GaAs/Al$_{0.3}$Ga$_{0.7}$As square quantum
well, \cite{PRL,quantumwell} without any externally applied static electric and magnetic fields.
We make the simplifying assumption of parabolic conduction subbands
(i.e., neglecting the difference of the effective masses in well and barriers).
Furthermore, we neglect BIA, but assume spin splitting is dominated by SIA, 
described by a simplified Rashba term of the form \cite{rashba}
\begin{equation} \label{4.1}
 \hat{H}\sia = \alpha[\vec{\sigma}\times {\bf q}]_z  = 
 \left( \begin{array}{cc} 0 & R \\ R^* & 0 \end{array} \right),
 \end{equation}
where $R = \alpha(q_y + i q_x)$, and  $\alpha$ is taken to be a real, positive 
adjustable parameter. The Rashba field is thus assumed to be the same for all
conduction subbands, which is a reasonable approximation for wide quantum wells.
The two-component Kohn-Sham equation (\ref{II.4}) becomes
\begin{equation} \label{4.2}
\left( \begin{array}{cc} \hat{h}_0 + v^{\rm xc}_{\uparrow \uparrow} & R + v^{\rm xc}_{\uparrow \downarrow}\\
R^* + v^{\rm xc}_{\downarrow \uparrow} & \hat{h}_0 + v^{\rm xc}_{\downarrow \downarrow} \end{array}
\right)
\left( \begin{array}{c} \psi_{i\uparrow} \\ \psi_{i\downarrow} \end{array} \right) = 
E_{i \qqp} 
\left( \begin{array}{c} \psi_{i\uparrow} \\ \psi_{i\downarrow} \end{array} \right) ,
\end{equation}
where $ i = 1,2,3,\ldots$, and
\begin{equation} \label{4.3}
\hat{h}_0 = \frac{1}{2m^*} \left(-\frac{d^2}{dz^2} + \qp^2 \right)+ v_{\rm conf}(z) + v_{\rm H}(z) \;.
\end{equation}
Equation (\ref{4.2}) is solved by the following ansatz:
\begin{eqnarray}\label{4.4}
\psi_{sj\uparrow}(\qqp,z) &=& \frac{1}{\sqrt{2}} \: \varphi_j(z) \\
\psi_{sj\downarrow}(\qqp,z) &=& \frac{s}{\sqrt{2}} \: \frac{R^*}{|R|} \: \varphi_j(z)\;,
\label{4.5}
\end{eqnarray}
where we replaced the subband index $i$ by the pair of indices $\{sj\}$, such that
$s = (-1)^i$ and $j = (i+1)/2$ for $i$ odd and $j=i/2$ for $i$ even. In the absence of
the off-diagonal terms in Eq. (\ref{4.2}), i.e. for inversion symmetry and hence spin degeneracy
at each $\qqp$, 
$j$ simply labels the spin-degenerate pairs, and $s$ labels the members of the pairs.

It is not difficult to see that in the presence of $R$ the ground-state density matrix
remains diagonal with $\nuu = \ndd$, and hence
$v^{\rm xc}_{\uparrow \uparrow} = v^{\rm xc}_{\downarrow \downarrow} \equiv v_{\rm xc}$
and $v^{\rm xc}_{\uparrow \downarrow} = v^{\rm xc}_{\downarrow \uparrow} = 0$.
The $\varphi_j(z)$ are therefore simply the solutions of the spin-unpolarized effective-mass Kohn-Sham
equation 
\begin{equation}
\left[- \frac{1}{2m^*}\frac{d^2}{dz^2}+ v_{\rm conf} + v_{\rm H}+ v_{\rm xc} \right] 
\varphi_j = \epsilon_j \varphi_j \;,
\end{equation}
where $\epsilon_j$ are the energy levels of the associated, doubly degenerate, parabolic subbands.
The presence of the off-diagonal Rashba terms in Eq. (\ref{4.2}), however,  lifts the 
spin degeneracy for $\qqp \ne 0$.
We thus obtain, using $|R| = \alpha \qp$,
\begin{equation}
E_{sj \qqp} = \epsilon_j + \frac{\qp^2}{2m^*} + s \alpha \qp \;, \qquad s=\pm 1\;,
\end{equation}
for the energy eigenvalues associated with the solutions (\ref{4.4}),(\ref{4.5}) of Eq. (\ref{4.2}).

\subsection{Collective intersubband excitations}

In the following, we will consider only cases where the lowest conduction
band is occupied, which restricts the electron density in the quantum well
to $N<1.82 \times 10^{11}\: \rm cm^{-2}$.
The goal is to study collective charge- and spin-density excitations
between the first and the second subband.
These collective modes are obtained by solving the response equation (\ref{III.12}) for
the case where the external perturbation is zero. In that case, the effective perturbing
potential $V_k^{(1)}$ consists of the self-consistent linearized Hartree and xc terms only.

\begin{figure}
\unitlength1cm
\begin{picture}(5.0,6.3)
\put(-6.,-9.3){\makebox(5.0,6.3){
\includegraphics{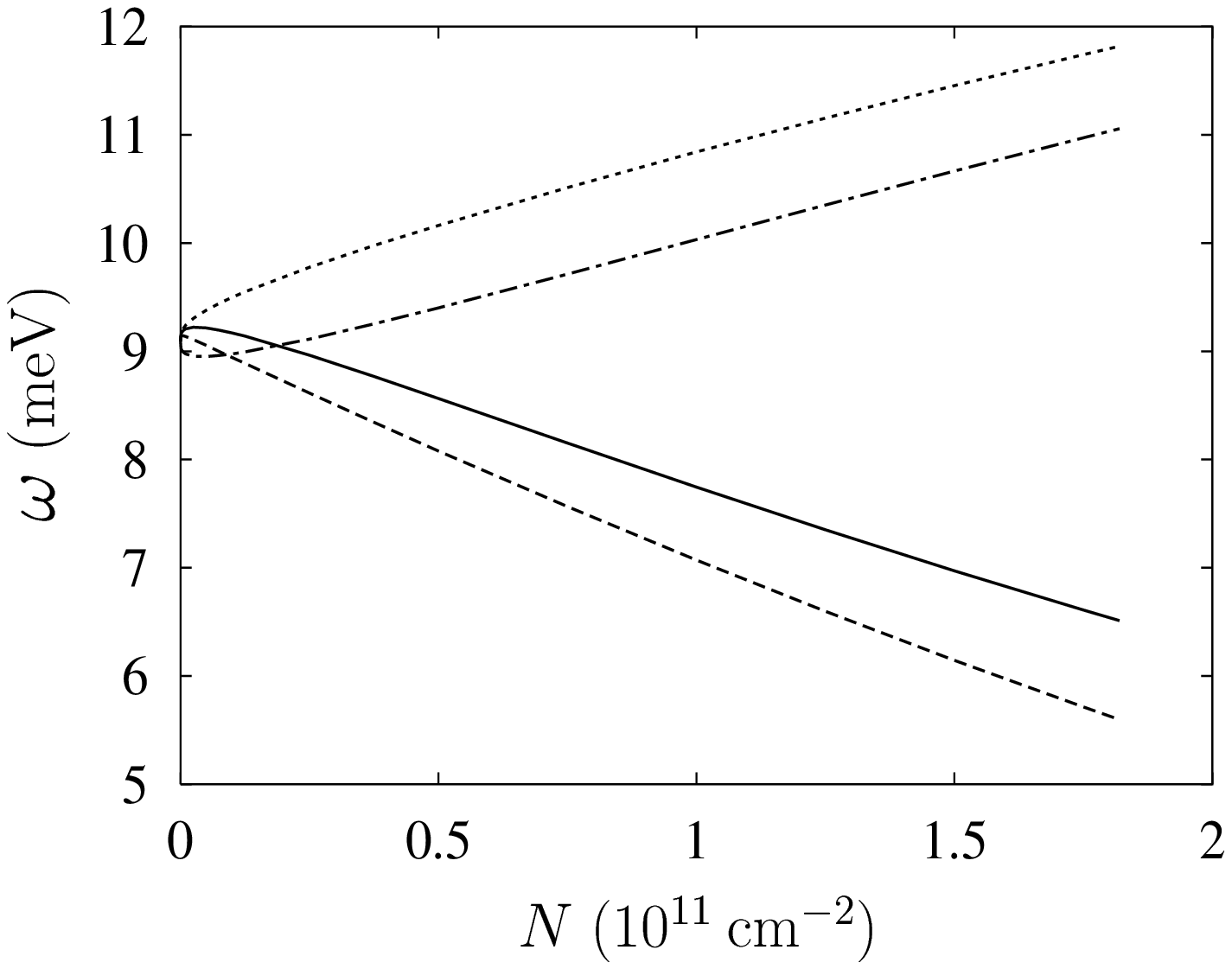}
}}
\end{picture}
\caption{Lowest ($1$$\to$$\,2$)
ISB excitation frequencies  versus electronic sheet density $N$ in a 40 nm GaAs/Al$_{0.3}$Ga$_{0.7}$As
quantum well, at $\kkp=0$,
for values of $N$ where only the lowest subband is occupied
(with parabolic subbands and without spin-orbit splitting).
Full line: single-particle excitations ($\omega=\epsilon_2 - \epsilon_1$).
Dotted line: charge-density excitations in RPA.
Dash-dotted and dashed lines: charge- and spin-density excitations in ALDA. 
}
\label{figure1}
\end{figure}

Let us first consider the case without spin-orbit coupling ($\alpha=0$).
Figure \ref{figure1} shows the density dependence of various ISB excitations at
zero in-plane wavevector ($\kkp=0$). The full line depicts the single-particle excitations 
with frequencies $\omega=\epsilon_2 - \epsilon_1$, i.e. the bare Kohn-Sham excitation energies.
The dotted line shows the ISB charge-density excitation in RPA, i.e. setting $f^{\rm xc}_{kl}=0$
in the effective potential $V_k^{(1)}$, Eq. (\ref{III.14}). The RPA excitation energies are
always higher than the single-particle excitations, due to the so-called depolarization shift.
\cite{vinter} The ISB charge-density excitation in ALDA is shown by the dash-dotted line.
Including xc effects in the response calculation produces a downshift of the plasmon 
energy of up to 0.75 meV. Finally, the spin-density ISB excitation is shown by the dashed
line. In RPA, this excitation coincides with the single-particle excitation, since the
depolarization shift affects only the charge mode.  Thus, the spin plasmon only exists as a distinct, 
collective excitation because of xc effects.

We now include spin-orbit coupling in the quantum well
material by taking a finite, density-independent value of $\alpha = 10$ meV{\AA} for the Rashba coupling
parameter. This is a typical value for practical situations of interest,
for instance when applying a static electric field of strength $10 \: \rm kV/cm$ in
a GaAs quantum well. \cite{andrada}

\begin{figure}
\unitlength1cm
\begin{picture}(5.0,11.5)
\put(-7.1,-9.9){\makebox(5.0,11.5){
\includegraphics{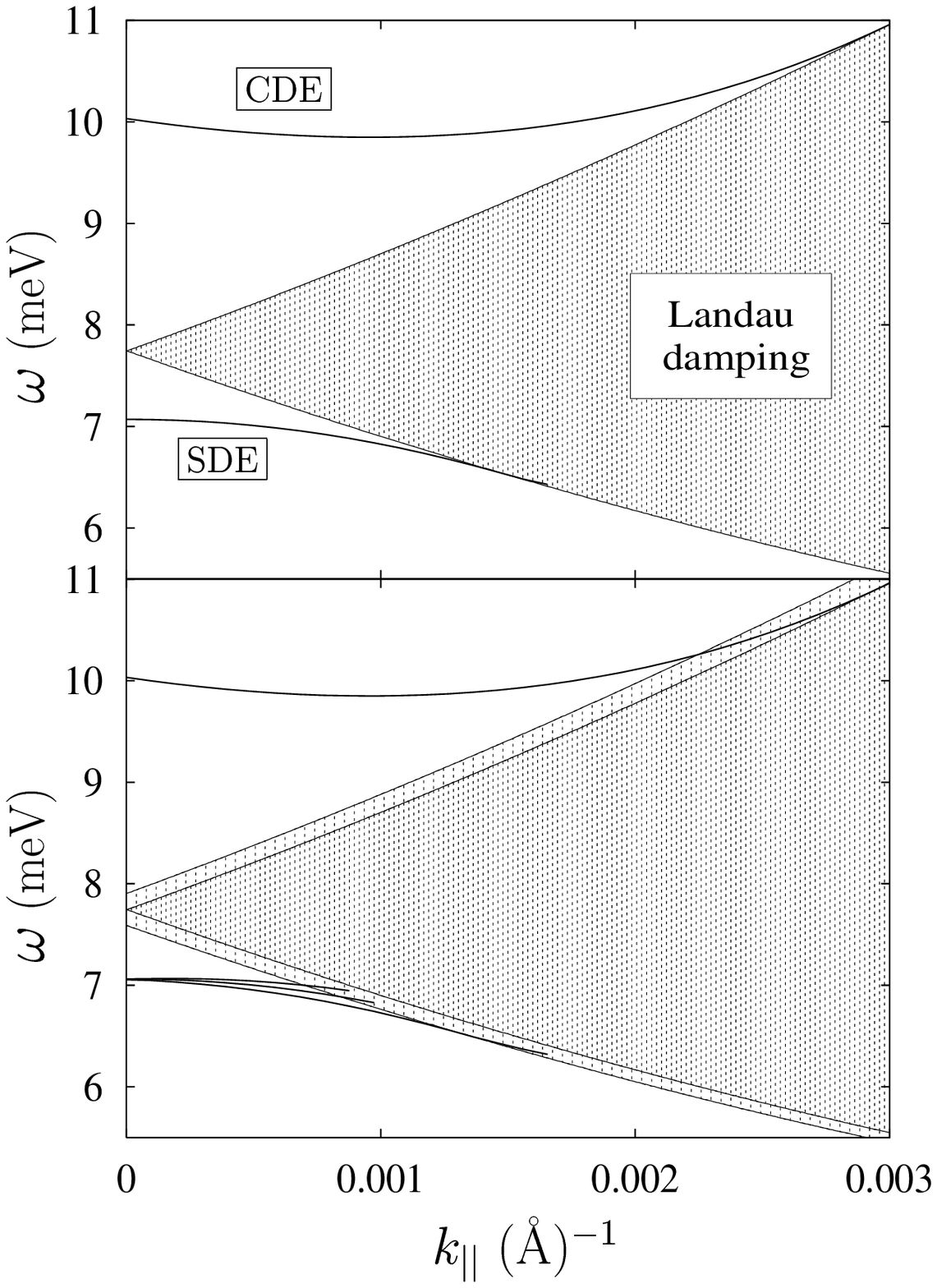}
}}
\end{picture}
\caption{
ISB charge and spin plasmon wavevector dispersions in
a 40 nm GaAs/Al$_{0.3}$Ga$_{0.7}$As quantum well,
for Rashba coefficients $\alpha=0$ (top) and  $\alpha = 10$ meV{\AA} (bottom).
The electronic sheet density is $1\times 10^{11}\: \rm cm^{-2}$.
The shaded regions indicate Landau damping of the charge and spin plasmons. For  $\alpha=0$, both
regions coincide. For finite $\alpha$, the Landau damping region for charge plasmons is unchanged (darker region),
but grows for spin plasmons (darker plus lighter region). The charge plasmon
is essentially independent of $\alpha$, but the spin plasmon splits into three branches for finite $\alpha$.
}
\label{figure2}
\end{figure}

The ALDA in-plane wavevector dispersions of the ISB plasmons are shown in Figure \ref{figure2},
comparing the case of $\alpha=0$ (top) and finite $\alpha$ (bottom). 
The shaded regions indicate Landau damping,
i.e., collective modes overlap with the particle-hole continuum and can decay into
incoherent particle-hole pairs. In both cases, the charge plasmon lies
above the region of Landau damping, and the spin
plasmons lie below. In the case of $\alpha=0$, there is a common region of Landau damping for the
charge- and the spin plasmons. For finite $\alpha$, the region of Landau damping
for the spin plasmons grows, while for the charge plasmons it stays unchanged. 
In the absence of other intrinsic or extrinsic scattering mechanisms (phonons, disorder),
all collective modes outside the region of Landau damping have infinite lifetime in ALDA. 

\begin{figure}
\unitlength1cm
\begin{picture}(5.0,6.5)
\put(-6.2,-9.3){\makebox(5.0,6.5){
\includegraphics{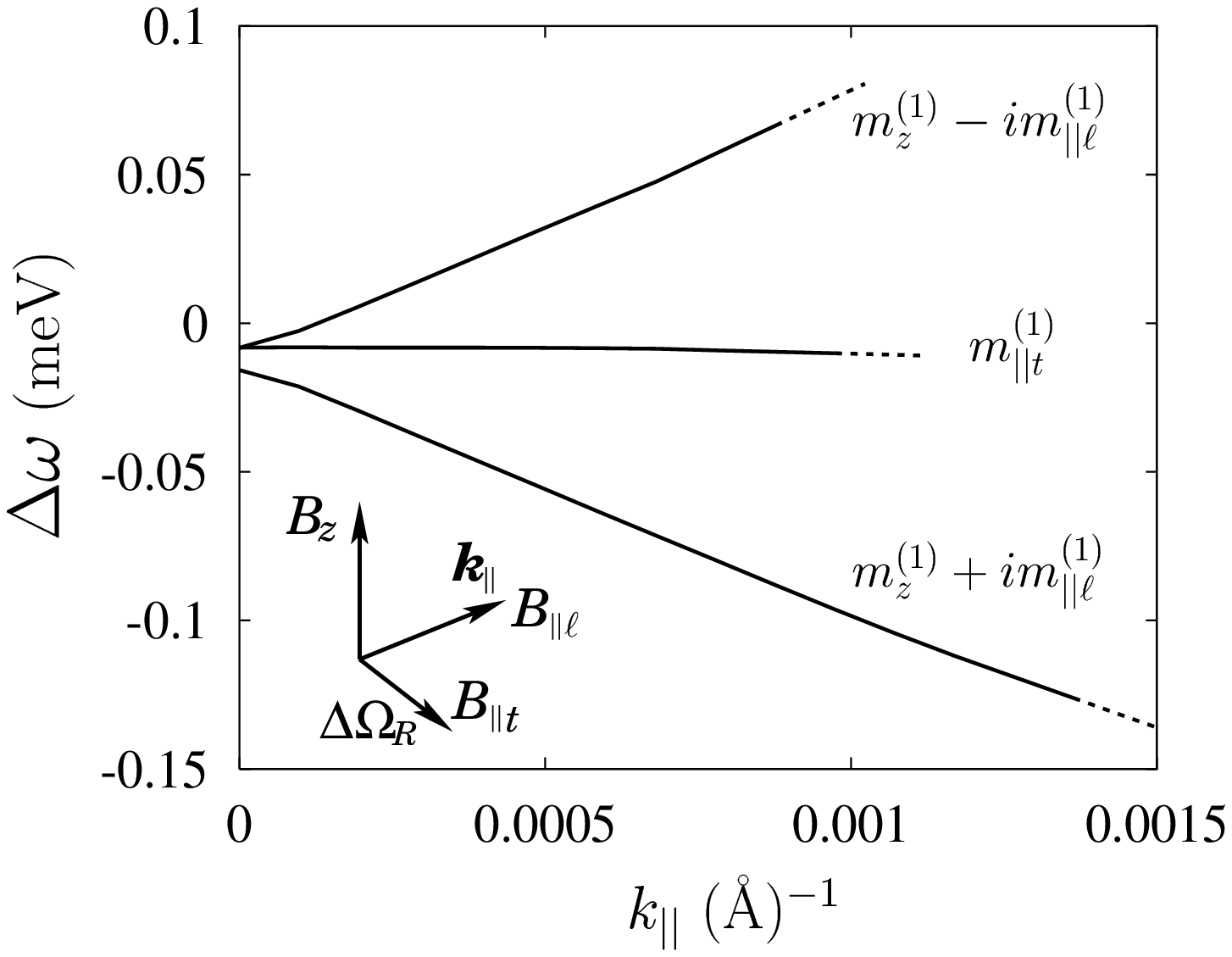}
}}
\end{picture}
\caption{ Splitting of the ISB spin plasmon dispersions, for the same quantum well
as in Figure \ref{figure2}. $\Delta \omega$ denotes the difference of the
spin plasmon frequencies at $\alpha = 10$ meV{\AA} and $\alpha=0$. The dots indicate
that the plasmons enter the region of Landau damping. The inset
illustrates the selection rules (see text). $m_z^{(1)}$ and $m_{||\ell}^{(1)}$ are coupled and two-fold
split. To lowest order in $\alpha$, the splitting has the form  $S=C(N)\alpha\kp$, where 
$C(N)$ depends on the electron sheet density.
}
\label{figure3}
\end{figure}

The charge plasmon dispersion is essentially independent of $\alpha$. The spin plasmon, however,
splits up into three branches for finite $\alpha$. This is shown in more detail in Figure \ref{figure3}, where
$\Delta \omega$ denotes the difference of the spin plasmon frequencies at $\alpha = 10$ meV{\AA} and
$\alpha=0$. There are three different spin plasmon modes, all degenerate at $\alpha=0$.
We will now discuss the nature of these modes, and how they couple to external fields.

The charge and spin plasmons with $\kkp= 0$ couple to external spin-dependent potentials
of the form $\underline{v}^{(1,\rm ext)}(z,\omega) = e E_0 z \sigma_j$.
For $j=0$ (CDE), $\underline{v}^{(1,\rm ext)}$
is related to an oscillating uniform  electric field 
${\bf E} \exp(-i\omega t)$, where ${\bf E} = E_0 \hat{e}_z$ (linearly
polarized along $z$, perpendicular to the quantum well plane). 
For the SDE's ($j=1,2,3$), $\underline{v}^{(1,\rm ext)}$ corresponds
by comparison with Eq. (\ref{II.7}) to oscillating magnetic fields 
${\bf B} \exp( - i\omega t)$, where $B = 2 E_0 z/g^*(z)\mu_B $.
The CDE and SDE's can thus be formally viewed as collective electric and magnetic dipole transitions.
At finite $\kkp$, the plasmons couple to external potentials of the form \cite{dobson} 
$\underline{v}^{(1,\rm ext)}(\kkp,z,\omega) = e E_0 a_0^* \exp(\kp z) \sigma_j$, where $a_0^*$ is the effective Bohr radius. 
As before, these potentials can be related to oscillating electric and
magnetic fields, ${\bf E}\exp[i(\kkp \cdot\rrp - \omega t)]$ and
${\bf B}\exp[i(\kkp \cdot\rrp - \omega t)]$.

The inset in Figure \ref{figure3} illustrates the selection rules for the individual SDE modes: 
(i) a longitudinal mode, denoted by $m_z^{(1)}$, which couples to a magnetic field
perpendicular to the quantum well, ${\bf B} = B_z \hat{e}_z$. 
(ii) two transverse (or spin-flip) modes, $m_{|| \ell}^{(1)}$ and $m_{||t}^{(1)}$, which couple 
to magnetic fields in the plane of the quantum well,
${\bf B} = B_{||\ell} \hat{e}_\ell$ and 
${\bf B} = B_{||t} \hat{e}_t$, where $\hat{e}_\ell = \kkp/\kp$ and
$\hat{e}_t = \hat{e}_\ell \times \hat{e}_z$.
Figure \ref{figure3} shows that, at finite $\alpha$, $m_z^{(1)}$ and $m_{||\ell}^{(1)}$
are coupled and two-fold split. On the other hand, $m_{||t}^{(1)}$
depends only very little on $\alpha$, except a small redshift independent of $\kp$. 
This small redshift, as well as the small splitting between the $z$ and $||\ell$ modes at $\kp=0$,
can be shown to be proportional to $\alpha^2$. 

Writing the Hamiltonian (\ref{4.1}) in the form $\hat{H}\sia = \hbar {\bf \Omega}_R \cdot \vec{\sigma}/2$ 
defines the Rashba effective magnetic field ${\bf \Omega}_R = (2\alpha/\hbar)(q_y,-q_x,0)$, which lies in the
quantum well plane and is perpendicular to $\qqp$. Since in our example all subbands experience the
same ${\bf \Omega}_R$, a collective ISB excitation with wavevector $\kkp$ implies a change in 
the effective magnetic field $\Delta {\bf \Omega}_R = (2\alpha/\hbar)(k_y,-k_x,0)$ for all single-particle
transitions, where $\Delta {\bf \Omega}_R || \hat{e}_t$ (see Figure \ref{figure3}). 
This explains the physical origin of the splitting between the different SDE branches:
The two spin plasmon branches whose energies are shifted ($z$ and $||\ell$) are those responding to fields
perpendicular to $\Delta {\bf \Omega}_R$, whereas the one which to lowest order in $\alpha$ does not shift 
($||t$) is parallel to $\Delta {\bf \Omega}_R$. Thus a spin polarization in either $z$ or $||\ell$
will precess in the $z-\ell$ plane. 
There are two possible linear combinations, $m_z^{(1)} \pm i m_{||\ell}^{(1)}$,
one precessing in that direction which is favored by $\Delta {\bf \Omega}_R$, the other in the opposite
direction, thus costing more energy.

\begin{figure}
\unitlength1cm
\begin{picture}(5.0,6.)
\put(-5.9,-9.){\makebox(5.0,6.){
\includegraphics{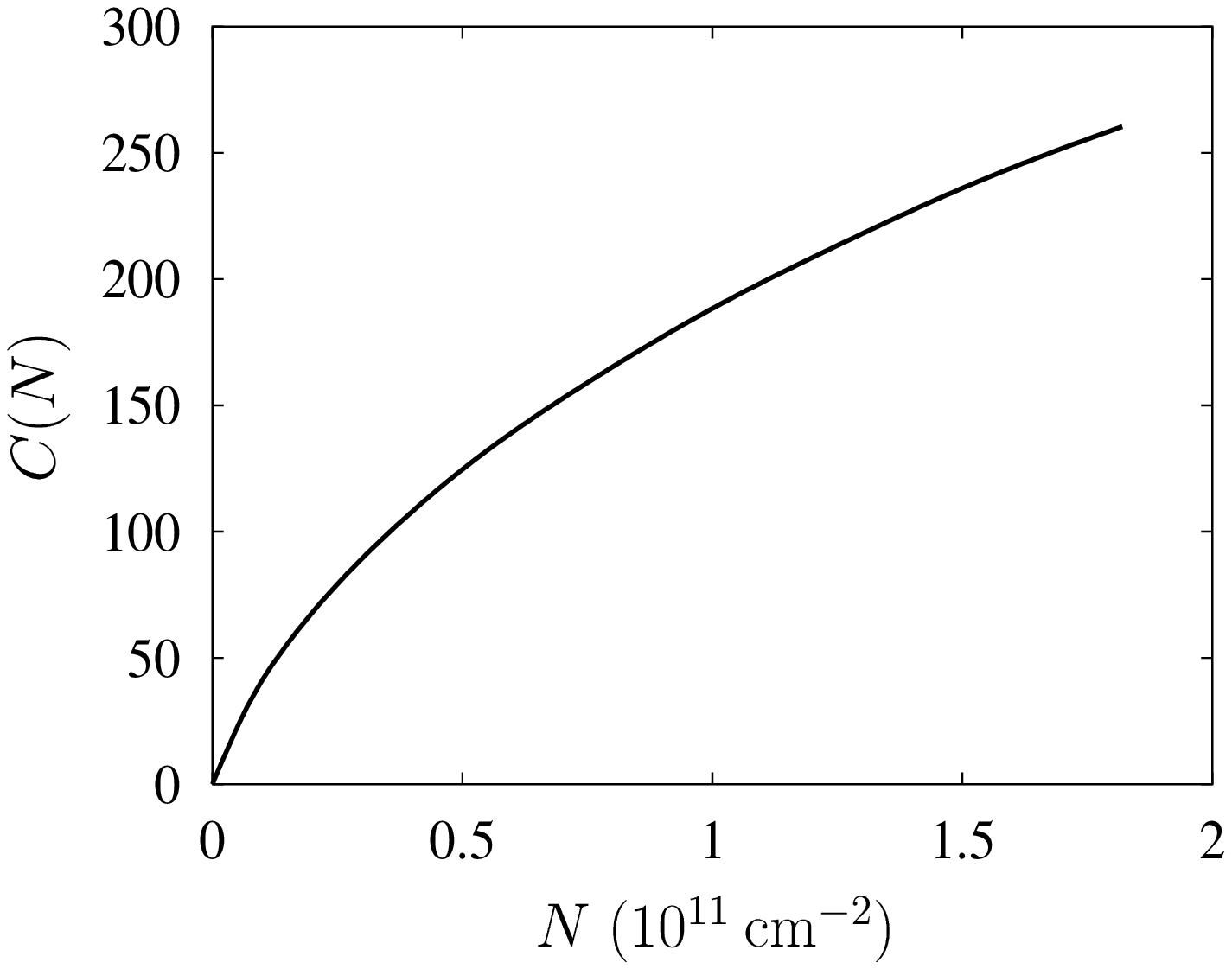}
}}
\end{picture}
\caption{
ISB  spin plasmon splitting coefficient $C(N)$ (see Figure \ref{figure3}), versus electron sheet density.}
\label{figure4}
\end{figure}

To lowest order in $\alpha$,
the magnitude of the splitting  between the two linear combinations of $m_z^{(1)}$ and $m_{||\ell}^{(1)}$, $S$,
is  proportional to $\alpha$ and grows linearly with $\kp$, i.e. $S = C(N) \alpha \kp$. The numerical prefactor $C(N)$ is a 
function of the electron density, and is plotted in Figure \ref{figure4}. The
SDE splitting strongly increases with electron density, and reaches values
of around $0.1 - 0.2$ meV for sheet densities of order $1\times 10^{11} \: \rm cm^{-2}$ and higher,
and plasmon wavevectors of order $0.001$ {\AA}$^{-1}$.
These splittings should be experimentally observable, for example using inelastic light scattering
techniques. \cite{abstreiter,pinczuk} This would provide an opportunity for measuring the
Rashba coefficient $\alpha$. 

%************************************************************************************
\section{conclusion} \label{sec:conclusion}
We have presented a microscopic theory of collective charge- and spin-density
excitations in  semiconductor quantum wells based on spin-density-functional theory,
with specific emphasis on intersubband excitations within the conduction band.
The approach consists
of two steps. We first calculate the electronic ground state in the quantum well
(subband levels and envelope functions), including conduction band non-parabolicity and spin
splitting, which leads to a $2\times 2$ conduction band Hamiltonian. The associated Kohn-Sham
matrix equation features spin-dependent xc potentials which are functionals of the spin-density matrix.

We then determine the excitation energies using linear response theory in the formulation
of TDDFT. Formally, one needs to solve a $2\times 2$ matrix equation for the coupled charge-
and spin-density-matrix response, including dynamic many-body effects.

To illustrate the formalism, we considered the example of a quantum well with
parabolic subbands that are split by a Rashba effective magnetic field. The charge plasmons were
found to be independent of the Rashba field. The three possible spin plasmons, which
are degenerate in the absence of spin-orbit coupling, were found to be split into
three branches, the splitting being proportional to the in-plane wavevector and to the strength
of the Rashba field.

This study illustrates the importance of including many-body effects beyond the RPA
in calculating collective spin excitations. The collective nature of the ISB spin plasmons is purely a consequence
of dynamical xc effects. Due to these collective effects, $T_2^{\scy ISB}$ is {\em not} 
influenced by the precessional decoherence mechanisms related to spin-orbit coupling 
which determine the intraband spin relaxation time $T_2$. \cite{DP}
Therefore, in the absence of impurities, disorder and phonon scattering,
the lifetime of ISB spin plasmons is limited by dynamical many-body effects only.
To capture these effects, one has to go beyond the ALDA and include retardation. \cite{PRL}
Within the ALDA, on the other hand, collective CDE and SDE's are infinitely long-lived.

The effect of non-parabolic bands and more general forms of spin-orbit splitting (both BIA and SIA) 
in semiconductor quantum wells will be addressed in more quantitative detail in the future.

%************************************************************************************
\begin{acknowledgments}
This work was supported in part by DARPA/ARO DAAD19-01-1-0490
and by the University of Missouri Research Board.
\end{acknowledgments}
%****************************************************************************************
\appendix
\section{LSDA for noncollinear spins}
\label{appendixA}

In LSDA,\cite{gunnarsson} the xc energy per particle of a spin-polarized homogeneous electron gas,
$\exc(n,\xi)$, is usually approximated by the von Barth--Hedin para\-me\-tri\-za\-tion:
\cite{hedin,dreizlergross} 
\begin{equation} \label{a1}
\exc(n,\xi) = \exc(n,0) + [\exc(n,1)-\exc(n,0)] f(\xi).
\end{equation}
The interpolation function between the paramagnetic ($\xi=0$) and ferromagnetic ($\xi=1$)
limits,
\begin{equation} \label{a2}
f(\xi) = \frac{(1+\xi)^{4/3} + (1-\xi)^{4/3} -2}{2(2^{1/3}-1)} \; ,
\end{equation}
reproduces the exact $\xi$-dependence of the exchange energy of the homogeneous electron gas,
and approximates the $\xi$-dependence of the correlation energy.

In the LSDA for non-collinear spins,\cite{gunnarsson,kubler,sandratskii,heinonen}
one still uses the spin-polarized homogeneous electron gas as reference
system, assuming that the xc energy per particle depends only on the local ground-state density $n$ and the
absolute value of the local ground-state spin polarization $\vec{\xi}$, where, using definitions
(\ref{II.18}) and (\ref{II.19}),
\begin{equation}\label{a3}
n = n_{\uparrow \uparrow} + n_{\downarrow \downarrow} 
\end{equation}
\begin{equation}\label{a4}
\vec{\xi} =  
\frac{1}{n}\left( \begin{array}{c} n_{\uparrow\downarrow} + n_{\downarrow \uparrow} \\
i(n_{\uparrow \downarrow} - n_{\downarrow \uparrow}) \\
n_{\uparrow \uparrow} - n_{\downarrow \downarrow} \end{array} \right)
\equiv
\frac{1}{n} \left( \begin{array}{c} m_1 \\ m_2 \\ m_3 \end{array}\right)
\end{equation}
so that
\begin{equation} \label{a5}
|\xi|= \frac{1}{n}\:\sqrt{ m_1^2 + m_2^2 + m_3^2} \;,
\end{equation}
and the sign of $\xi$ is determined with respect to the chosen global 
quantization axis: $\mbox{sign}(\xi) = \mbox{sign}(\nuu-\ndd)$. 

The xc potential in LSDA may then be obtained as follows:
Letting $n \equiv m_0$, we define
\begin{equation}
v^{\rm xc}_j(z) = \left.\frac{\partial[n\exc(n,\xi)]}{\partial m_j} \right|
_{\scriptstyle m_i = m_i(z) \atop \scriptstyle i=0,1,2,3}
\end{equation}
which yields
\begin{eqnarray}\label{a6}
v^{\rm xc}_0 &=& 
\exc + n \frac{\partial \exc}{\partial n} - \xi \frac{\partial \exc}{\partial \xi} \\
v^{\rm xc}_{i} &=& \frac{m_i}{n\xi} \frac{\partial
\exc}{\partial \xi} \;, \qquad i=1,2,3\:.
\label{a7}
\end{eqnarray}
Using
\begin{equation}\label{a8}
v^{\rm xc}_{\alpha \beta} = \sum_{i=0}^3 \frac{\partial m_i}{\partial n_{\alpha \beta}}
\: v^{\rm xc}_{i} \;, \qquad \alpha,\beta=\uparrow,\downarrow\;,
\end{equation}
one finds
\begin{equation}\label{a9}
v^{\rm xc}_{\uparrow \uparrow} = e_{\rm xc} + n\: \frac{ \partial e_{\rm xc}}{\partial n}
+ \left[ \frac{\nuu - \ndd}{n \zeta}- \zeta \right]
\frac{\partial e_{\rm xc}}{\partial \zeta}
\end{equation}

\begin{equation}\label{a10}
v^{\rm xc}_{\downarrow \downarrow}= e_{\rm xc}+ n\: \frac{ \partial e_{\rm xc}}{\partial n}
-  \left[ \frac{\nuu - \ndd}{n \zeta}+ \zeta \right]
\frac{\partial e_{\rm xc}}{\partial \zeta}
\end{equation}

\begin{equation}\label{a11}
v^{\rm xc}_{\uparrow \downarrow} = 
\frac{2 \ndu}{n \zeta}\: \frac{\partial e_{\rm xc}}{\partial \zeta}
\end{equation}

\begin{equation}\label{a12}
v^{\rm xc}_{\downarrow \uparrow} = 
\frac{2 \nud}{n \zeta}\: \frac{\partial e_{\rm xc}}{\partial \zeta} \;.
\end{equation}
Equations (\ref{a9})--(\ref{a12}) are in agreement with the results of Heinonen {\em et al.} \cite{heinonen}

Next, we calculate, in ALDA, the xc kernels needed in Equation (\ref{III.14}). The definition is
\begin{equation} \label{a13}
f^{\rm xc}_{jk}(z,z',\omega) =\left.
\frac{\partial^2 [n \exc(n,\xi)]}{\partial m_j \partial m_{k}}
 \: \right|_{\scriptstyle m_i = m_i(z) \atop \scriptstyle i=0,1,2,3 }
\delta(z-z') \;,
\end{equation}
with the following results [omitting the $\delta(z-z')$]:
\begin{equation}\label{a14}
f^{\rm xc}_{00} = 2 \frac{\partial \exc}{\partial n} + n \frac{\partial^2 \exc}{\partial n^2}
- 2 \xi \frac{\partial^2 \exc}{\partial n \partial \xi}
+ \frac{ \xi^2}{n}\: \frac{\partial ^2 \exc}{\partial \xi^2} \label{fxcnn} 
\end{equation}

\begin{equation}\label{a15}
f^{\rm xc}_{0 i} = 
\frac{m_i}{n \xi}\: \frac{\partial^2 \exc}{\partial n \partial \xi}
- \frac{m_i}{n^2} \: \frac{\partial ^2 \exc}{\partial \xi^2} 
\end{equation}

\begin{equation}\label{a16}
f^{\rm xc}_{ij}  =
\frac{\delta_{ij}}{n \xi} \: \frac{\partial \exc}{\partial \xi}
- \frac{m_i m_j}{(n\xi)^3} \: \left( \frac{ \partial \exc}{\partial \xi}
- \xi \: \frac{\partial^2 \exc}{\partial \xi^2} \right) ,
\end{equation}
where $i,j=1,2,3$ in (\ref{a15}) and (\ref{a16}), and $f^{\rm xc}_{ij} = f^{\rm xc}_{ji} $
for all $i,j$. For spin unpolarized ground states ($m_1 = m_2 = m_3= 0$),
only those xc kernels diagonal in $i,j$ are nonzero, with
\begin{equation}\label{a17}
f^{\rm xc}_{00}  = 2 \frac{\partial \exc(n,0)}{\partial n} + n \frac{\partial^2 \exc(n,0)}{\partial n^2}
\end{equation}
and
\begin{equation} \label{a18}
f^{\rm xc}_{ii} = \frac{4/9}{n(2^{1/3}-1)} \: [\exc(n,1) - \exc(n,0)] 
\end{equation}
for $i=1,2,3$. Notice that $f^{\rm xc}_{00}$ and $f^{\rm xc}_{ii}$ have the same exchange parts,
$f^{\rm x}_{00} = f^{\rm x}_{ii} =
(4/9n)e_{\rm x}^h(n,0) = -(9 \pi n^2)^{-1/3}$, but in general have different
correlation parts.

%********************************************************************************************
\section{Non-interacting response functions}\label{appendixB}
For convenience, we list here the explicit relations between the 
response functions $\Pi\ks_{jk}$ and $\chi\ks_{\sigma\sigma',\lambda\lambda'}$
following from Equations (\ref{III.10}), (\ref{III.11}) and (\ref{III.13}) (omitting
the superscript ``KS''):
\begin{eqnarray*}
\Pi_{00} &=& \chi_{\uparrow \uparrow,\uparrow \uparrow} + 
\chi_{\uparrow \uparrow,\downarrow \downarrow} +
\chi_{\downarrow \downarrow,\uparrow \uparrow} +
\chi_{\downarrow \downarrow,\downarrow \downarrow} 
\\
\Pi_{01} &=& \chi_{\uparrow \uparrow,\uparrow \downarrow} + 
\chi_{\uparrow \uparrow,\downarrow \uparrow} +
\chi_{\downarrow \downarrow,\uparrow \downarrow} +
\chi_{\downarrow \downarrow,\downarrow \uparrow} 
\\
\Pi_{02} &=& -i(\chi_{\uparrow \uparrow,\uparrow \downarrow} - 
\chi_{\uparrow \uparrow,\downarrow \uparrow} +
\chi_{\downarrow \downarrow,\uparrow \downarrow} -
\chi_{\downarrow \downarrow,\downarrow \uparrow}) 
\\
\Pi_{03} &=& \chi_{\uparrow \uparrow,\uparrow \uparrow} -
\chi_{\uparrow \uparrow,\downarrow \downarrow} +
\chi_{\downarrow \downarrow,\uparrow \uparrow} -
\chi_{\downarrow \downarrow,\downarrow \downarrow}
\end{eqnarray*}
\begin{eqnarray*}
\Pi_{10} &=& \chi_{\uparrow \downarrow,\uparrow \uparrow} + 
\chi_{\uparrow \downarrow,\downarrow \downarrow} +
\chi_{\downarrow \uparrow,\uparrow \uparrow} +
\chi_{\downarrow \uparrow,\downarrow \downarrow}
\\
\Pi_{11} &=& \chi_{\uparrow \downarrow,\uparrow \downarrow} + 
\chi_{\uparrow \downarrow,\downarrow \uparrow} +
\chi_{\downarrow \uparrow,\uparrow \downarrow} +
\chi_{\downarrow \uparrow,\downarrow \uparrow}
\\
\Pi_{12} &=& -i(\chi_{\uparrow \downarrow,\uparrow \downarrow} - 
\chi_{\uparrow \downarrow,\downarrow \uparrow} +
\chi_{\downarrow \uparrow,\uparrow \downarrow} -
\chi_{\downarrow \uparrow,\downarrow \uparrow})
\\
\Pi_{13} &=& \chi_{\uparrow \downarrow,\uparrow \uparrow} -
\chi_{\uparrow \downarrow,\downarrow \downarrow} +
\chi_{\downarrow \uparrow,\uparrow \uparrow} -
\chi_{\downarrow \uparrow,\downarrow \downarrow}
\end{eqnarray*}
\begin{eqnarray*}
\Pi_{20} &=& i(\chi_{\uparrow \downarrow,\uparrow \uparrow} + 
\chi_{\uparrow \downarrow,\downarrow \downarrow} -
\chi_{\downarrow \uparrow,\uparrow \uparrow} -
\chi_{\downarrow \uparrow,\downarrow \downarrow})
\\
\Pi_{21} &=& i(\chi_{\uparrow \downarrow,\uparrow \downarrow} + 
\chi_{\uparrow \downarrow,\downarrow \uparrow} -
\chi_{\downarrow \uparrow,\uparrow \downarrow} -
\chi_{\downarrow \uparrow,\downarrow \uparrow})
\\
\Pi_{22} &=& \chi_{\uparrow \downarrow,\uparrow \downarrow} - 
\chi_{\uparrow \downarrow,\downarrow \uparrow} -
\chi_{\downarrow \uparrow,\uparrow \downarrow} +
\chi_{\downarrow \uparrow,\downarrow \uparrow}
\\
\Pi_{23} &=& i(\chi_{\uparrow \downarrow,\uparrow \uparrow} -
\chi_{\uparrow \downarrow,\downarrow \downarrow} -
\chi_{\downarrow \uparrow,\uparrow \uparrow} +
\chi_{\downarrow \uparrow,\downarrow \downarrow})
\end{eqnarray*}
\begin{eqnarray}
\Pi_{30} &=& \chi_{\uparrow \uparrow,\uparrow \uparrow} + 
\chi_{\uparrow \uparrow,\downarrow \downarrow} -
\chi_{\downarrow \downarrow,\uparrow \uparrow} -
\chi_{\downarrow \downarrow,\downarrow \downarrow} 
\nonumber\\
\Pi_{31} &=& \chi_{\uparrow \uparrow,\uparrow \downarrow} + 
\chi_{\uparrow \uparrow,\downarrow \uparrow} -
\chi_{\downarrow \downarrow,\uparrow \downarrow} -
\chi_{\downarrow \downarrow,\downarrow \uparrow} 
\nonumber\\
\Pi_{32} &=& -i(\chi_{\uparrow \uparrow,\uparrow \downarrow} - 
\chi_{\uparrow \uparrow,\downarrow \uparrow} -
\chi_{\downarrow \downarrow,\uparrow \downarrow} +
\chi_{\downarrow \downarrow,\downarrow \uparrow}) 
\nonumber\\
\Pi_{33} &=& \chi_{\uparrow \uparrow,\uparrow \uparrow} -
\chi_{\uparrow \uparrow,\downarrow \downarrow} -
\chi_{\downarrow \downarrow,\uparrow \uparrow} +
\chi_{\downarrow \downarrow,\downarrow \downarrow} \;.
\label{b1}
\end{eqnarray}
With eigenfunctions of the form (\ref{4.4}),(\ref{4.5}), the Kohn-Sham response function (\ref{III.9})
can be written as
\begin{eqnarray}\label{b2}
\chi\ks_{\sigma \sigma',\lambda\lambda'}(\kkp,z,z',\omega) &=& \sum_{ij} 
F^{ij}_{\sigma\sigma',\lambda\lambda'}(\kkp,\omega) \varphi_i(z) \varphi_j(z)    \nonumber\\
&&\times  \varphi_i(z') \varphi_j(z') \;,
\end{eqnarray}
and likewise
\begin{equation}\label{b4}
\Pi\ks_{kl}(\kkp,z,z',\omega) = \sum_{ij} 
G^{ij}_{kl}(\kkp,\omega) \varphi_i(z) \varphi_j(z)    
 \varphi_i(z') \varphi_j(z') .
\end{equation}
The $G^{ij}_{kl}(\kkp,\omega)$ are related to the
$F^{ij}_{\sigma\sigma',\lambda \lambda'}(\kkp,\omega)$ according to (\ref{b1}).
The latter functions are given by
\begin{widetext}
\begin{eqnarray}
F^{ij}_{\sigma\sigma',\lambda \lambda'}(\kkp,\omega) &=&
-\frac{1}{4}\sum_{ss'}^{\pm 1} \! \int\!\frac{d^2\qp}{(2 \pi)^2}  \:
\frac{f(E_{si\qqp}) }
{\omega - E_{s i \qqp} + E_{s'j\qqp-\kkp}+ i\eta}
\left[ \delta_{\sigma\uparrow}
+ \delta_{\sigma\downarrow} \: s \: \frac{R^*(\qqp)}{|R(\qqp)|} \right]
\nonumber\\
&&  \times
\left[ \delta_{\sigma'\uparrow}
+ \delta_{\sigma'\downarrow} \: s' \: \frac{R(\qqp-\kkp)}{|R(\qqp-\kkp)|} \right]
\left[ \delta_{\lambda\uparrow}
+ \delta_{\lambda\downarrow} \: s \: \frac{R(\qqp)}{|R(\qqp)|} \right]
\left[ \delta_{\lambda'\uparrow}
+ \delta_{\lambda'\downarrow} \: s' \: \frac{R^*(\qqp-\kkp)}{|R(\qqp-\kkp)|} \right]
\nonumber\\[3mm]
&+&\frac{1}{4}
\sum_{pp'}^{\pm1}\! \int\! \frac{d^2\qp}{(2 \pi)^2} \:
\frac{f(E_{si\qqp}) }
{\omega + E_{s i \qqp} - E_{s'j\qqp+\kkp}+ i\eta} 
\left[ \delta_{\sigma\uparrow}
+ \delta_{\sigma\downarrow} \: s' \: \frac{R^*(\qqp+\kkp)}{|R(\qqp+\kkp)|} \right]
\nonumber\\
&& \times
\left[ \delta_{\sigma'\uparrow}
+ \delta_{\sigma'\downarrow} \: s \: \frac{R(\qqp)}{|R(\qqp)|} \right]
\left[ \delta_{\lambda\uparrow}
+ \delta_{\lambda\downarrow} \: s' \: \frac{R(\qqp+\kkp)}{|R(\qqp+\kkp)|} \right]
\left[ \delta_{\lambda'\uparrow}
+ \delta_{\lambda'\downarrow} \: s \: \frac{R^*(\qqp)}{|R(\qqp)|} \right] .
\label{fij}
\end{eqnarray}
It is not difficult to show that for the case of $\alpha\to 0$, i.e. vanishing Rashba
term, the function $F^{ij}_{\sigma \sigma',\lambda \lambda'}$ becomes diagonal in the spin indices and reduces to the 
well-known limit\cite{quantumwell} ($\omega_{ij} = \epsilon_j - \epsilon_i$)
\begin{equation}
F^{ij}_{\sigma\sigma',\lambda \lambda'}(\kkp,\omega) =
- \delta_{\sigma\lambda} \delta_{\sigma' \lambda'}\int\!\frac{d^2\qp}{(2 \pi)^2}\left\{
\frac{f(\epsilon_i + \frac{\qp^2}{2m^*}) }
{\omega + \omega_{ij} + \frac{\kp^2}{2m^*} - \frac{\qqp\cdot \kkp}{m^*} + i\eta}
-
\frac{f(\epsilon_i + \frac{\qp^2}{2m^*}) }
{\omega - \omega_{ij} - \frac{\kp^2}{2m^*} - \frac{\qqp \cdot \kkp}{m^*}+ i\eta}
\right\}.
\end{equation}
\end{widetext}

%***************************************************************************
%***************************************************************************
%***************************************************************************

\end{document}